\documentclass[twocolumn,,amssymb,amsmath,aps,showpacs,floatfix,nofootinbib,showpacs,balancelastpage]{revtex4}
\usepackage{amsmath}
\usepackage{amssymb}

\usepackage{hangcaption,fancybox,feynmp}
\usepackage{indentfirst}
\usepackage{graphicx}
\usepackage{epsfig,subfigure,psfrag}
\usepackage{mathrsfs}

\begin{document}


\title{A `fast-burning' mechanism for magnetic diffusion}
\author{
Bo Xiao$^{1}$\footnote{E-mail:homenature@pku.edu.cn},
Zhuo-wei Gu$^{1}$,
Ming-xian Kan$^{1}$,
Gang-hua Wang$^{1}$\footnote{E-mail:wanggh@caep.cn},
and Jian-heng Zhao$^{1}$ 
}
\affiliation{ $ ^1$ Institute of Fluid Physics, CAEP, P. O. Box 919-105, Mianyang 621900, China }

\date{\today}

\begin{abstract}

Fast-burning mechanism describes the rapid penetration, with a sharp-shaped wave-front, 
of a strong magnetic field into a conductive metal whose electric resistance poses an 
abrupt rise at some critical temperature.
With its wave-front sweeping over a solid metal, the fast-burning
can melt or vaporize the metal very rapidly.
This paper derives formulas for the existence conditions and wave-front velocity of a fast-burning.


\end{abstract}


\maketitle

\section{Introduction}\label{sec.introduction}

In many strong magnetically driven experiments such as magnetically driven flyer plate, magnetically driven line implosion,
and explosive magnetic flux compression, it is important for the metal media to keep in solid state as long as possible.
One reason is that solid state of the metal can reduce the growth rate \cite{Robert2001, Atchison1997} of the magneto-Rayleigh-Taylor (MRT) instability
\cite{Harris1962, Lau2011, Sinars2010, Sinars2011} which breaks up the integrity of the system 
and is one of the main threat to the success of those experiments \cite{Almstrom1997, Sinars2010, Sinars2011}.
Another reason, for magnetically driven flyer plate experiment \cite{Sun2014}, 
is that the back-end of the flyer plate must keep in solid state for being able to be implemeted in impact experiments.

\begin{figure}[htpb]
\centering
  \includegraphics[width=2.5in]{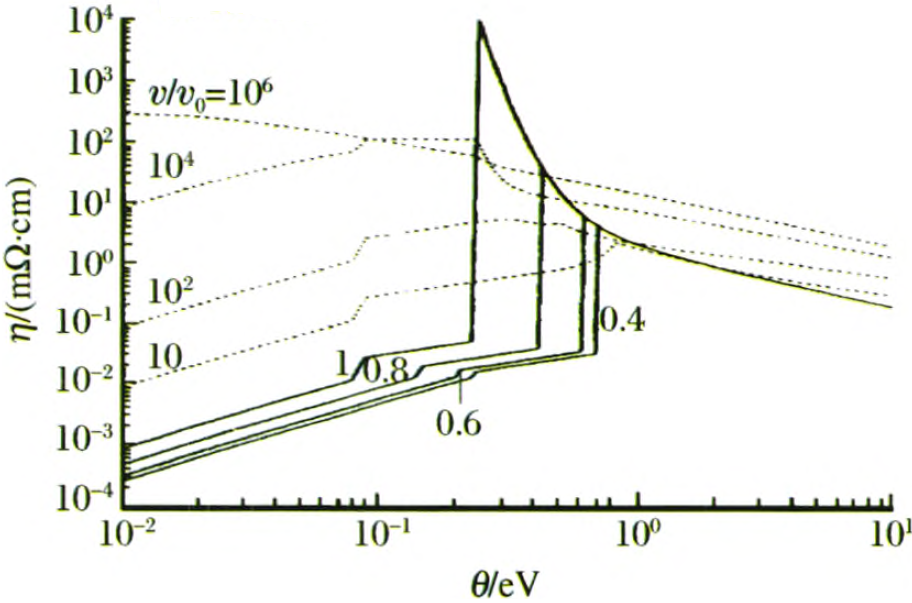}\\
  \caption{Temperature dependence of electric resistance of Aluminum. \cite{Kan2013,Burgess1986} }\label{fig.ResistanceOfAl}
\end{figure}

The electric resistance of most metals increases with temperature in the condensed region,
and especially, the resistance usually poses a jump at some critical temperature near the vaporization point.
A typical example of the temperature dependence of metal resistance is shown in Fig. \ref{fig.ResistanceOfAl}.

Considering a general magnetically driven experiment that
a solid metal media is pushed by a strong magnetic field,
shortly after the time started, the magnetic field diffuses into some depth of the metal,
and Joule heat is generated together and causes a rise of temperature in the diffused region of the metal.
The increase of temperature causes an increase of the metal's resistance 
due to the positive dependence of resistance on temperature, 
which in return accelerates the magnetic diffusion and causes the Joule heat to be generated faster. 
Those positive feedbacks acting iteratively may leads finally 
to a fast penetration of the magnetic field into the metal in case some conditions are satisfied.
Since the critical temperature is usually near the vaporization point, 
this fast penetration would melt or vaporize the metal in a short time, 
thus we call this phenomenon `fast-burning'.

The fast-burning phenomenon may have appeared in many magneto-hydrodynamics simulations 
of strong magnetically driven experiments where temperature dependence of resistance is considered \cite{Peterson2012, Peterson2014}.
However, untill now, no specialized study of fast-burning are found in literatures. 
In this paper, we have a theoretical study of fast-burning, 
deriving formulas for its existence conditions and penetrating velocity.

\section{Fast-burning solvation of magnetic diffusion equation}\label{sec.analytic}

\subsection{Physical problem modeling}\label{seq.problemModeling}

\begin{figure}[htpb]
\centering
  \subfigure[]{\includegraphics[width=2.0in]{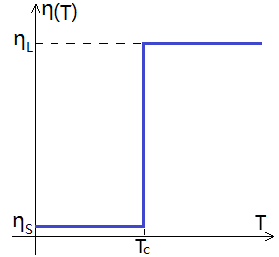}\label{fig.EtaToT}}\\
  \subfigure[]{\includegraphics[width=2.0in]{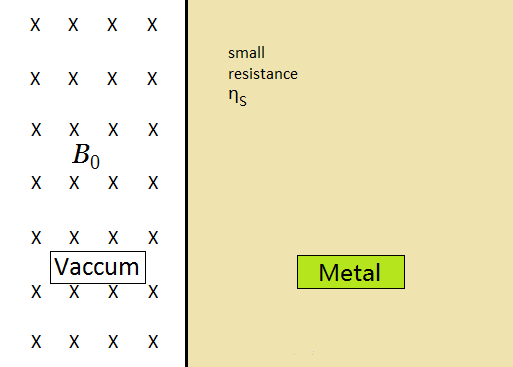}\label{fig.FastBurnIllustrate1}}\\
  \subfigure[]{\includegraphics[width=2.0in]{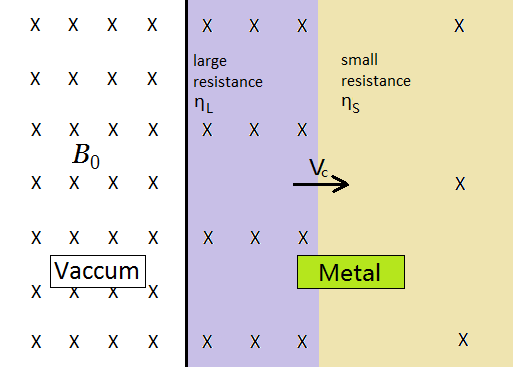}\label{fig.FastBurnIllustrate2}}\\
  \caption{(a): The metal's electric resistance $\eta$ evolving with temperature $T$, with $\eta_{\rm L} >> \eta_{\rm S}$. 
           (b): At the initial state, a metal slab is exposed to a vacuum magnetic field. 
           (c): A fast-burning is propagating in the metal after the time started.}\label{fig.PhysicalProblem}
\end{figure}

Let's first establish a simplified mathematical description of the physical problem to be discussed.
The physical problem is a vacuum magnetic field diffuses into a metal slab, as illustrated in Fig. \ref{fig.PhysicalProblem}.
Kinematic movement is not considered, and temperature diffusion, whose rate is usually $10^{-3}$ smaller than magnetic diffusion, 
is also neglected. Only magnetic diffusion is considered in the system, i.e., the control equation is solely
\begin{equation}\label{eq.contronEquation}
  \frac{\partial}{\partial t} B(x,t) = -\frac{\eta}{\mu_0} \frac{\partial^2}{\partial x^2} B(x,t)
\end{equation}
The metal's electric resistance poses a typical jump at a critical temperature $T_{\rm c}$,
as modeled in Fig. \ref{fig.EtaToT}, where $\eta_{\rm L} \gg \eta_{\rm S}$ is assumed.
The vacuum magnetic field $B_0$ is kept constant in the process.
At the initial state, the magnetic field distribution in the metal slab is zero.

After the time started, the vacuum magnetic field diffuses into the metal,
and Joule heat is generated in companion.
If the Joule heat density is large enough, it would rise the metal's temperature above
its critical temperature $T_{\rm c}$, and cause an abrupt increase of its resistance.
The increase of resistance will accelerate the magnetic field diffusion.
Those effects act iteratively will cause a fast penetration of the magnetic field into the metal,
as illustrated in Fig. \ref{fig.FastBurnIllustrate2}. 
In the following, for convenience of discussion, we also talk about the critical Joule heat density $J_{\rm c}$,
which is defined as the Joule heat per volume (not per mass) that is required to heat the metal to its critical temperature.

\subsection{Fast-burning solvation of magnetic diffusion equation}\label{seq.fastBurningSolvation}

\begin{figure}[htpb]
\centering
  \subfigure[]{\includegraphics[width=2.5in]{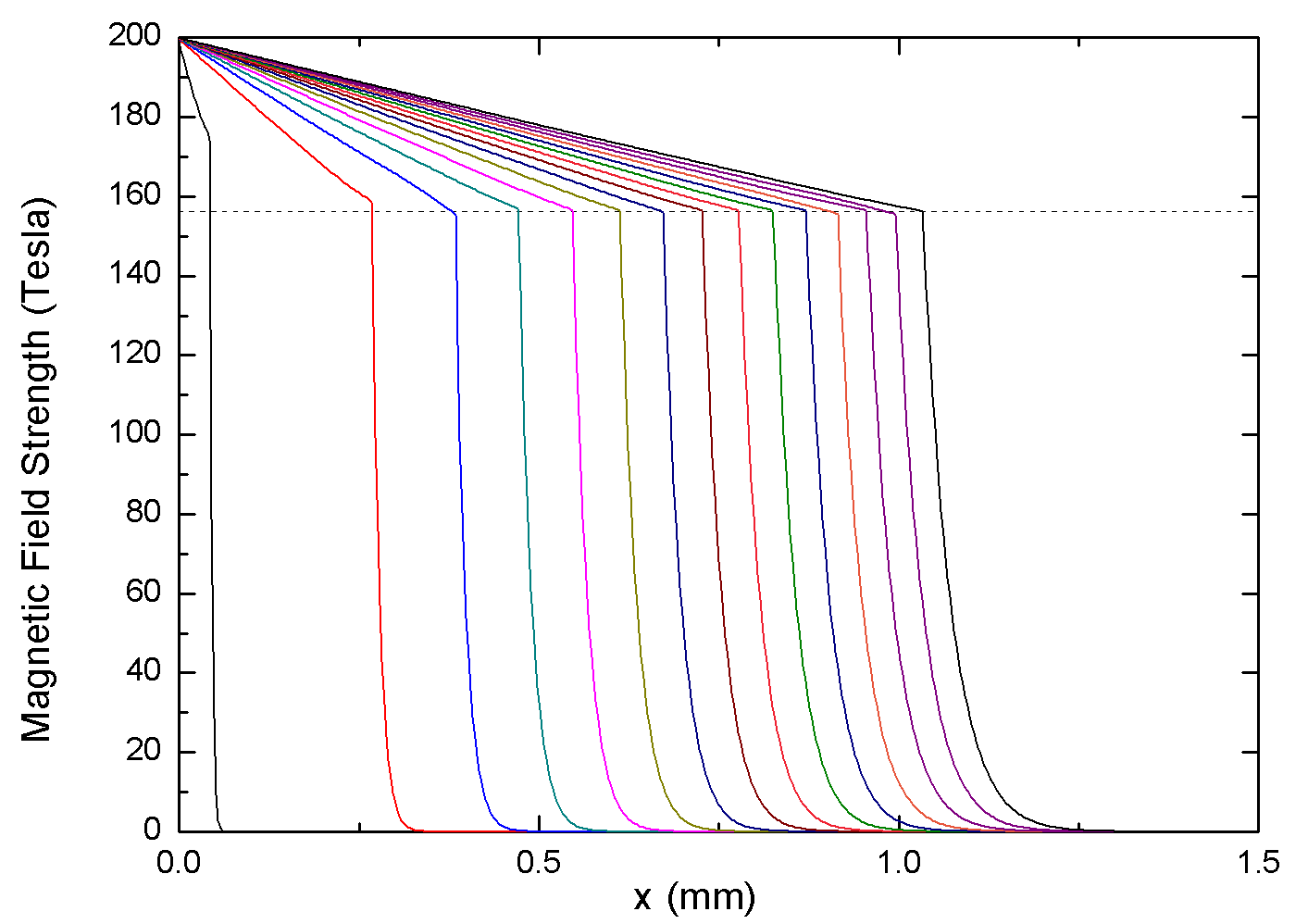}\label{fig.BzEvolveLinear}}\\
  \subfigure[]{\includegraphics[width=2.5in]{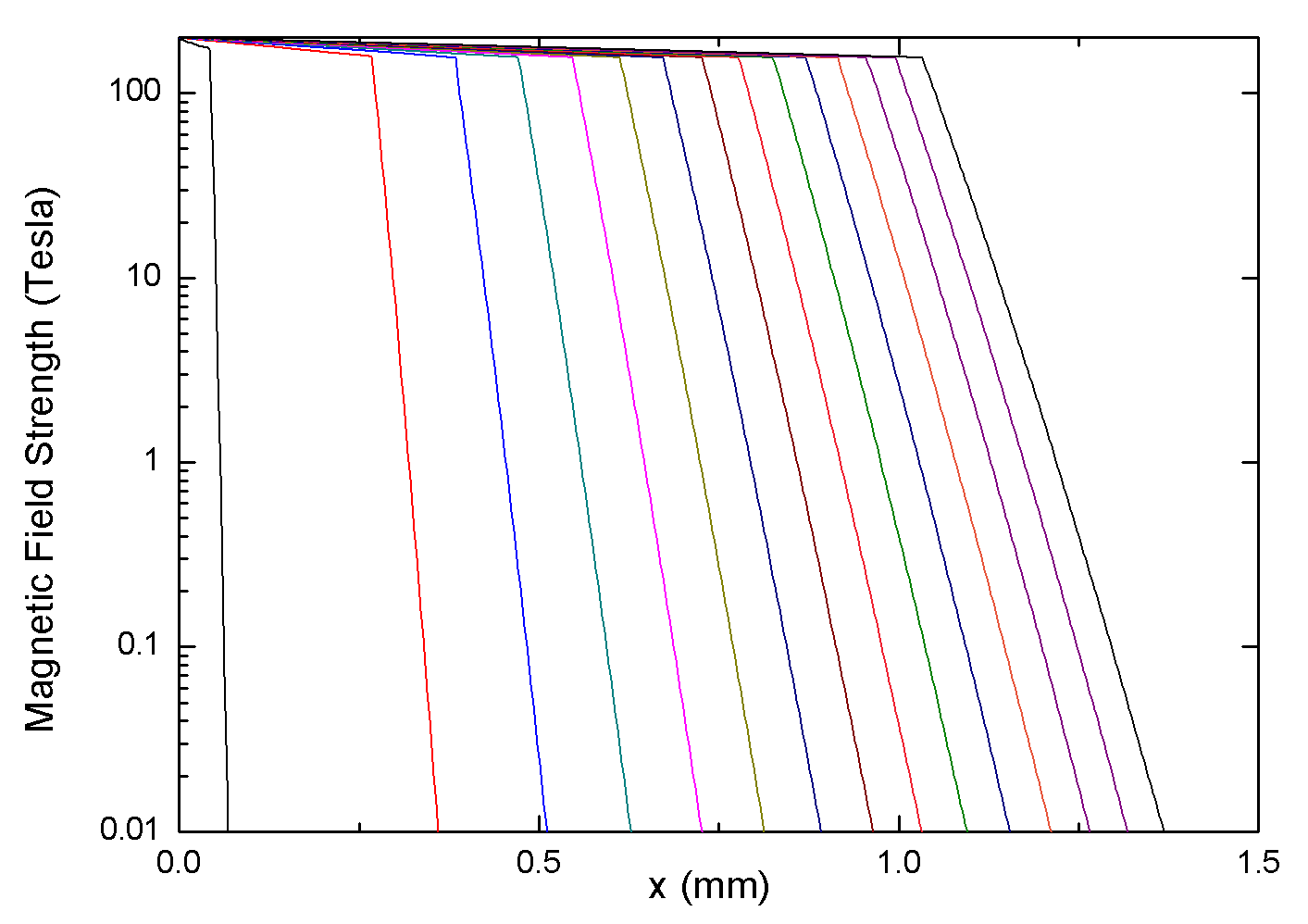}\label{fig.BzEvolveLog}}\\
  \caption{Evolution of the magnetic field distribution with time. 
           Figure (b) shows the same results of (a) in Log coordinate.
           Each curve in the figures represents a magnetic field distribution at a different time, 
           with a time increase of $0.02 \mu s$ between neighbored lines from left to right. 
           A dashed horizontal line is drawn in figure (a) to mark the height of the knees.
  }\label{fig.BzEvolve}
\end{figure}

First, a one-dimensional numerical simulation of the physical problem of Fig. \ref{fig.PhysicalProblem} was carried.
In the simulation, the metal's parameters are set according to the material of steel.
The metal's critical Joule heat density is set $J_{\rm c} = 1.0\times 10^4{\rm J\;cm^{-3}}$
(which is estimated with the formula $J_{\rm c} = (T_{\rm c}-T_0) c_v \rho$,
with critical temperature $T_{\rm c}=3134{\rm K}$, room temperature $T_0=300{\rm K}$, 
heat capacity $c_v=0.45{\rm J\;g^{-1}K^{-1}}$, and mass density $\rho = 7.86{\rm g\;cm^{-3}}$).
The metal's resistance below $T_{\rm c}$ is set $\eta_{\rm S} = 9.7\times10^{-8}{\rm \Omega m}$
while its resistance above $T_{\rm c}$ is set $\eta_{\rm L} = 100\times\eta_{\rm S}$,
and the vacuum magnetic field strength is set $B_0 = 200{\rm T}$.
Simulation results of the time evolution of magnetic field distribution
are shown in Fig. \ref{fig.BzEvolve}. Three typical characteristics are shown in
the figure: 1. there is a `knee' on each distribution line, and the magnetic field strength
at the knees is nearly constant; 2. the distribution behind (leftside) the knee is very close to a linear distribution;
3. the distribution ahead (rightside) the knee is very close to an exponent.

\begin{figure}[htpb]
\centering
  \includegraphics[width=2.5in]{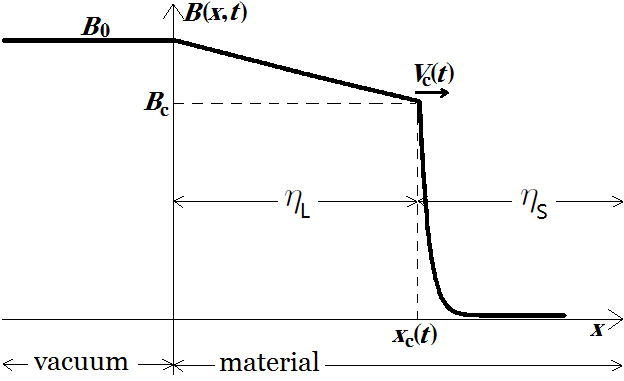}\\
  \caption{Magnetic field distribution of a fast-burning solvation of magnetic diffusion equation.}\label{fig.MagnetoDistribution}
\end{figure}

Inspired by the above numerical simulation, we guess a fast-burning solvation of the
magnetic diffusion equation as illustrated in Fig. \ref{fig.MagnetoDistribution}.
The solvation is departed into three sections: a constant field in the vacuum section,
a linear distribution in the large resistance $\eta_{\rm L}$ section,
and an exponent distribution in the small resistance $\eta_{\rm S}$ section, that is
\begin{subequations}\label{eq.fastBurningSolvation}
\begin{equation}
  B(x,t) = B_0, \texttt{for $x<0$}, \label{eq.constSection}\\
\end{equation}
\begin{equation}
  B(x,t) = B_0 - \frac{B_0-B_{\rm c}}{x_{\rm c}} x, \texttt{for $0<x<x_{\rm c}$}, \label{eq.linearSection}\\
\end{equation}
\begin{equation}
  B(x,t) = B_{\rm c} \exp \{ -\frac{\mu_0}{\eta_{\rm S}}V_{\rm c}(x-x_{\rm c})\}, \texttt{for $x_{\rm c} < x$}, \label{eq.exponentSection}\\
\end{equation}
\end{subequations}
where $x_{\rm c}$ is time dependent, $V_{\rm c}\equiv\frac{dx_{\rm c}}{dt}$, 
$B_{\rm c}$ corresponds to the height of the knees in Fig. \ref{fig.BzEvolveLinear} and is assumed a constant, 
and $B_0>B_{\rm c}$ is also assumed.

This solvation satisfy the magnetic diffusion equation (\ref{eq.contronEquation})
in an approximate sense under proper parameters configurations.
The exponent part Eq. (\ref{eq.exponentSection}) satisfies
$\frac{\partial}{\partial t}B = \frac{\eta_{\rm S}}{\mu_0}\frac{\partial^2}{\partial x^2}B$ approximately if
we have
\begin{equation}\label{eq.dVdtCondition0}
  \left| \frac{d V_{\rm c}}{d t} \right|(x-x_{\rm c}) \ll V_{\rm c}^2.
\end{equation}
The scale of $(x-x_{\rm c})$ can be estimated by the coefficient factor in the exponent, 
i.e. $(x-x_{\rm c})\sim \frac{\eta_{\rm S}}{\mu_0}\frac{1}{V_{\rm c}}$,
then the requirement (\ref{eq.dVdtCondition0}) becomes
\begin{equation}\label{eq.dVdtCondition}
  \left| \frac{d V_{\rm c}}{d t} \right|\frac{1}{V_{\rm c}^3}\frac{\eta_{\rm S}}{\mu_0} \ll 1.
\end{equation}
The linear part Eq. (\ref{eq.linearSection}) satisfy
$(\frac{\tilde{x}^2}{B_{\rm c}}\frac{\mu_0}{\eta_{\rm L}})\frac{\partial}{\partial t}B = (\frac{\tilde{x}^2}{B_{\rm c}})\frac{\partial^2}{\partial x^2}B$
approximately if we have this equation's left hand approaching zero, i.e.,
\begin{equation}\label{eq.quasiStaticCondition0}
  \left(\frac{\tilde{x}^2}{B_{\rm c}}\frac{\mu_0}{\eta_{\rm L}}\right)(B_0-B_{\rm c})\frac{1}{x_{\rm c}}V_{\rm c} \ll 1.
\end{equation}
The $\tilde{x}$ is a typical length scale used for length normalization and can also be set as $\tilde{x}\sim \frac{\eta_{\rm S}}{\mu_0}\frac{1}{V_{\rm c}}$,
then the requirement Eq. (\ref{eq.quasiStaticCondition0}) becomes
\begin{equation}\label{eq.quasiStaticCondition}
  \frac{\eta_s^2}{\mu_0\eta_{\rm L}}\frac{1}{V_{\rm c} x_{\rm c}}\frac{B_0-B_{\rm c}}{B_{\rm c}} \ll 1.
\end{equation}

$B_{\rm c}$ in the solvation Eq. (\ref{eq.fastBurningSolvation}) needs to be determined.
It is related to the critical temperature $T_{\rm c}$ defined in Fig. \ref{fig.EtaToT}.
At the position $x_{\rm c}$, the metal's resistance changes from $\eta_{\rm S}$ to $\eta_{\rm L}$, this means that
the Joule heat density produced at this point is right $J_{\rm c}$. This Joule heat is collected by the moving exponent section. 
So we have
\begin{equation}\label{eq.BcDetermine}
\begin{split}
  J_{\rm c} &=  \int_0^\infty \eta_{\rm S} j^2(x') \frac{dx'}{V(x')}\\
  &= \frac{1}{V_{\rm c}}\int_0^\infty \eta_{\rm S} \left(\frac{1}{\mu_0}\frac{\partial}{\partial x'} \left(B_{\rm c} \exp \{ -\frac{\mu_0}{\eta_{\rm S}}V_{\rm c} x'\}\right)\right)^2 dx',
\end{split}
\end{equation}
where $V(x')$ is the moving velocity of the exponent section $\exp \{ -\frac{\mu_0}{\eta_{\rm S}}V_{\rm c} x'\}$ 
when the relative position $x'$ of the exponent section is sweeping over the considered fixed point in the metal, 
and we have used an approximation $V(x')=V_{\rm c}$
which means that the velocity of the exponent section keeps nearly constant when its \emph{main body} is
sweeping over the considered fixed point.
From Eq. (\ref{eq.BcDetermine}) we get
\begin{equation}\label{eq.BcGot}
  B_{\rm c} = \sqrt{2\mu_0 J_{\rm c}}.
\end{equation}
Eq. (\ref{eq.BcGot}) also indicates an existence condition for fast-burning, that is, 
for a fast-burning solvation to exist, we must have
\begin{equation}\label{eq.ExistCondition}
  B_0 > \sqrt{2\mu_0 J_{\rm c}}.
\end{equation}

Velocity of fast-burning $V_{\rm c}(t)$ can be determined from energy conservation.
The magnetic energy flux at the point $x_{\rm c}$
should equal the sum of the magnetic energy increase rate and Joule heat production ratio at the right side of $x_{\rm c}$,
that is
\begin{equation}\label{eq.VcDetermine}
  -\frac{1}{\mu_0}E(x_{\rm c}^-)\times B_{\rm c}) = (\frac{1}{2\mu_0}B_{\rm c}^2)V_{\rm c} + \int_0^\infty \eta_{\rm S} j^2(x') dx',
\end{equation}
where $E(x_{\rm c}^-) = \frac{\eta_{\rm L}}{\mu_0}\frac{\partial B(x_{\rm c}^-)}{\partial x} = -\frac{\eta_{\rm L}}{\mu_0}\frac{B_{\rm c}-B_0}{x_{\rm c}}$, 
and the term $(\frac{1}{2\mu_0}B_{\rm c}^2)$ is the magnetic energy density at $x_{\rm c}$.
From Eq. (\ref{eq.VcDetermine}) we get
\begin{equation}\label{eq.VcGot}
  V_{\rm c} = \frac{\eta_{\rm L}}{\mu_0}\frac{B_0-B_{\rm c}}{B_{\rm c}}\frac{1}{x_{\rm c}}.
\end{equation}

By applying the formulas (\ref{eq.BcGot}) and (\ref{eq.VcGot}),
the requirements (\ref{eq.dVdtCondition}) and (\ref{eq.quasiStaticCondition}) can be rewrote respectively as
\begin{equation}
  \eta_{\rm L}/\eta_{\rm S} \gg 1
\end{equation}
and
\begin{equation}
  \frac{\eta_{\rm L}}{\eta_{\rm S}}\frac{B_0-B_{\rm c}}{B_{\rm c}} \gg 1
\end{equation}

\subsection{Verification of the formulas}\label{sec.Verification}

\begin{figure}[htpb]
\centering
  \includegraphics[width=2.5in]{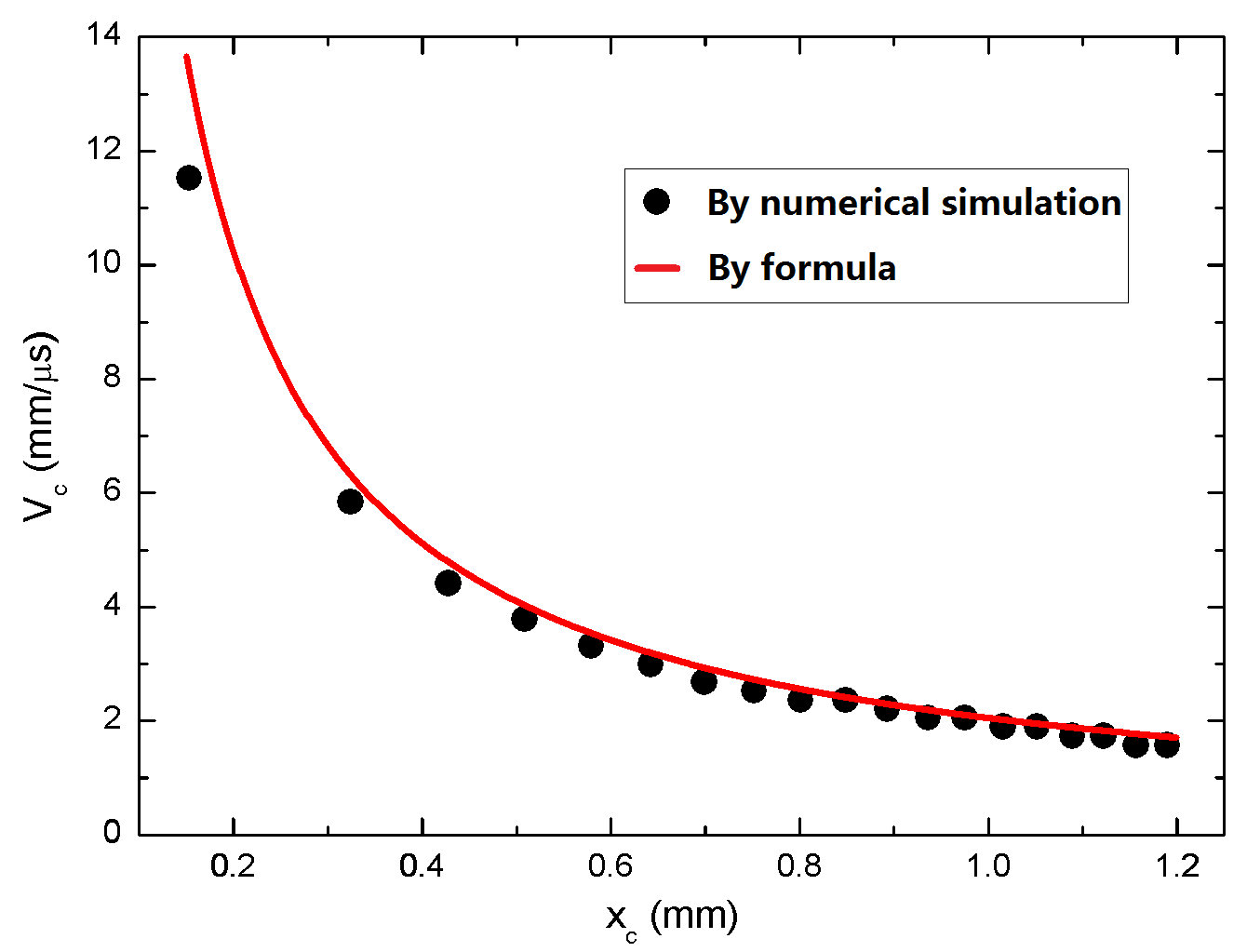}\\
  \caption{Comparison between formula and numerical simulation results of $V_{\rm c}$ under different $x_c$.}\label{fig.VcXc}
\end{figure}

The formulas for $B_{\rm c}$ and $V_{\rm c}$, i.e. Eq. (\ref{eq.BcGot}) and (\ref{eq.VcGot}), 
can be verified by comparing them with the one dimensional numerical simulations.
Using the same parameters as in the one dimensional simulation in section \ref{seq.fastBurningSolvation},
we obtain from formulas Eq. (\ref{eq.BcGot}) and (\ref{eq.VcGot}) the results
$B_{\rm c} = 1.58\times 10^2{\rm T}$ and $V_{\rm c} = (2.05 \rm m^2s^{-1})/x_{\rm c}$.
This value of $B_{\rm c}$ agrees well with the one dimensional numerical simulation result,
i.e. the height of the knees (the horizontal line) in Fig. \ref{fig.BzEvolveLinear}.
The result $V_{\rm c} = (2.05 \rm m^2s^{-1})/x_{\rm c}$ is drawn in Fig. \ref{fig.VcXc} together with the
results from one dimsional numerical simulations, and good agreement is also found.

\section{Conclusions and discussions}\label{sec.conclusion}

The conditions for the emerging of a fast-burning when a vacuum magnetic field diffuses into a metal whose resistance
poses a typical jump at some critical temperature are
$B_0>B_{\rm c}$ with $B_{\rm c}=\sqrt{2\mu_0 J_{\rm c}}$, $\eta_{\rm L}/\eta_{\rm S} \gg 1$, and $\frac{\eta_{\rm L}}{\eta_{\rm S}}\frac{B_0-B_{\rm c}}{B_{\rm c}} \gg 1$,
where $B_0$ is the vacuum magnetic field strength, $J_{\rm c}$ is the Joule heat density required to heat
the metal to its critical temperature, and $\eta_{\rm S}$ ($\eta_{\rm L}$) is the metal's electrical resistance below (above) critical temperature.

The velocity of the wave-front of a fast-burning is 
$V_{\rm c} = \frac{\eta_{\rm L}}{\mu_0}\frac{B_0-B_{\rm c}}{B_{\rm c}}\frac{1}{x_{\rm c}}$,
which is proportional to the metal's resistance above critical temperature 
and propotional to the difference between the vacuum magnetic field strength and the critical magnetic field strength.
The velocity gradually slows down when the wave-front propagate deeper (i.e. $x_{\rm c}$ becomes larger) into the metal.

The formulas derived in this paper can be used for estimating the time 
the metal keeps in solid state when strong magnetic field are penetrating it, 
and should be useful for the design of many strong magnetically driven solid media experiments.

{\em Acknowledgements:} Bo Xiao would like to thank Cheng-wei Sun and Shu-chao Duan for useful discussions. 
This work is supported partly by the Foundation of China Academy of Engineering Physics (No. 2015B0201023).


\begin{thebibliography}{10}


\bibitem{Robert2001}
E.~R.~Robert {\em et al.},
\newblock IEEE Transactions on Plasma Science, Vol. 30, No. 5:1764 (2002)

\bibitem{Atchison1997}
W.~L. Atchison, R.~J. Faehl, and R.~E. Reinovsky,
\newblock 11th IEEE Pulsed Power Conference, Vol. 1:372-377 (1997)

\bibitem{Harris1962}
E.~G. Harris, 
\newblock Phys. Fluids, 5, 1057-1062, 1962

\bibitem{Lau2011}
Y.~Y. Lau {\em et al.},
\newblock Physical Review E {\bf 83}, 066405 (2011)

\bibitem{Sinars2010}
D.~B. Sinars {\em et al.},
\newblock Phys. Rev. Lett. {\bf 105}, 185001 (2010)

\bibitem{Sinars2011}
D. B. Sinars {\em et al.},
\newblock Physics of Plasmas {\bf 18}, 056301 (2011)

\bibitem{Almstrom1997}
H.~Almstrom, S.~M. Golberg, and M.~A. Liberman,
\newblock 11th IEEE Pulsed Power Conference, Vol. 2:1405-1410 (1997)

\bibitem{Sun2014}
Y.~B. Sun and A.~R. Piriz,
\newblock Physics of Plasmas {\bf 21}, 072708 (2014)

\bibitem{Kan2013}
M.-x. Kan, G.-h. Wang, H.-l. Zhao, and L. Xie,
\newblock Exlposion and Shock Waves, Vol. 33, No. 3:282-286, May 2013

\bibitem{Burgess1986}
T.~J. Burgess,
\newblock SAND86-1093C, 1986.

\bibitem{Peterson2012}
K.~J. Peterson {\em et al.},
\newblock Physics of Plasmas {\bf 19}, 092701 (2012)

\bibitem{Peterson2014}
K.~J. Peterson {\em et al.},
\newblock Phys. Rev. Lett. {\bf 112}, 135002 (2014)


\end{thebibliography}
\end{document}